\documentclass[onecolumn,secnumarabic,amssymb,aps, prd]{revtex4-1} 

\usepackage{algorithm}
\usepackage{algorithmic}
\usepackage{graphicx} 
\usepackage{amsmath}

\begin{document}

\title{A Generative Model for Exploring Structure Regularities in Attributed Networks}%

\author{Zhenhai Chang}%
\email{changzhenhai2012@163.com}
\affiliation{1 School of Mathematics and Statistics, Tianshui Normal University, Gansu, China}
\affiliation{4 School of Statistics and Mathematics, Central University of Finance and Economics}

\author{Caiyan Jia}%
\email{cyjia@bjtu.edu.cn}
\affiliation{2 School of Computer and Information Technology, Beijing Jiaotong University, Beijing, China}
\affiliation{3 Beijing Key Lab of Traffic Data Analysis and Mining, Beijing, China}

\author{Xianjun Yin}%
\affiliation{4 School of Statistics and Mathematics, Central University of Finance and Economics}

\author{Yimei Zheng}
\affiliation{2 School of Computer and Information Technology, Beijing Jiaotong University, Beijing, China}
\affiliation{3 Beijing Key Lab of Traffic Data Analysis and Mining, Beijing, China}


\begin{abstract}
Many real-world networks known as attributed networks contain two types of information: topology information and node attributes. It is a challenging task on how to use these two types of information to explore structural regularities. In this paper, by characterizing potential relationship between link communities and node attributes, a principled statistical model named PSB\_PG that generates link topology and node attributes is proposed. This model for generating links is based on the stochastic blockmodels following a Poisson distribution. Therefore, it is capable of detecting a wide range of network structures including community structures, bipartite structures and other mixture structures. The model for generating node attributes assumes that node attributes are high dimensional and sparse and also follow a Poisson distribution. This makes the model be uniform and the model parameters can be directly estimated by expectation-maximization (EM) algorithm. Experimental results on artificial networks and real networks containing various structures have shown that the proposed model PSB\_PG is not only competitive with the state-of-the-art models, but also provides good semantic interpretation for each community via the learned relationship between the community and its related attributes. 
\end{abstract}

\maketitle
\tableofcontents

\section{Introduction}
Many complex systems in the real world take the form of networks, in which a collection of nodes joined together in pairs by edges or links. Examples include social networks, biological networks, and information networks [1]. One of the most important tasks in network analysis is to reveal community structures, where communities are groups of nodes with relatively dense connections within groups but sparse connections between them [2-4]. Besides, as emergence of online user-generated media (e.g., Twitter, Facebook and Microblogs), networks are not only characterized by graphs containing node connectivity, but each node also contains rich attribute information. It has attracted a lot of attention on how to identify community structures in these attributed networks (also called attributed graphs) in recent years [5-7]. In this situation, three ways can be used to detect communities: using attribute information only, using topological information only, combining the two types of information. Obviously, using only one type of information will ignore another type of information. It has shown that combing network topology with attribute information can not only improve the quality of community detection, but also has potential to provide the semantic descriptions of communities, and help to understand the functions of communities [7-11].

Existing methods that joint the two types of information can be roughly classified into two categories: model-based methods [12-21] and other heuristic methods [22-27]. Model-based methods are mainly on the basis of probabilistic generative models. They model the relationships between node attributes and network structures. By maximizing a corresponding joint likelihood function, the parameters including the node memberships and the relation between network structures and node attributes are inferred. 

In these models, some models uncover traditional community structure. For example, PCL\_DC [20] combined a popularity-based conditional link (PCL) model [20] for links and a discriminative content (DC) model [20] for node attributes. By introducing popularity and productivity of a node, PPL\_DC [19] extended PCL\_DC to directed and undirected networks. Since both PCL\_DC and PPL\_DC assumed that nodes from the same community were more likely to have links, the models could only detect traditional community structures also known as assortative structure [13].

Some models uncover a broad range of structures including assortative structure, disassortative structure [13] and other structures such as bipartite structures, core-peripheral structures and mixture structures. We call these broad range structures general structures for simplicity. The typical models of this kind are BNPA [12], PPSB\_DC [16], NEMBP [28]. BNPA combined Newman's mixture model (NMM) [29] and a multinomial distribution under the frame of Bayes to infer the number of communities and community structures. Thanks to NMM model, BNPA could detect general structures. PPSB\_DC made use of the advantages of PPL [19] that could capture various node degrees and GSB [30] that had the capability of detecting general structures. However, PPSB\_DC simply added the logistic DC model to the objective function of link model in the similar way with PPL\_DC. Therefore, it has no ability to describe the relationships between link communities and node attributes. In addition, we have found that the convergence of PPSB\_DC can not be guaranteed since it used a two-stage method to iteratively infer the model parameters. NEMBP combined degree-corrected stochastic blockmodel [31] and a multinomial distribution to detect structures, and had a good semantic interpretability because it characterized the relationship between a link community and its corresponding attribute cluster/topic. However, NEMBP needed to specify the number of topics and the number of communities in advance and supposed that the number of communities and the number of semantic topics were exact the same in all experiments. 

Bearing the convergence of a model and the ability to describing the diverse relationships between link communities and node attributes in mind, in this study, we propose a joint probabilistic generative model based on the assumption of block structure in stochastic blockmodels [32] and the idea of link communities [33-34] to generate both links and node attributes. In the model, the process for generating links and that for generating node attributes both follow Poisson distributions. This makes the likelihood function be constructed in an unified form and the model parameters can be directly estimated by expectation-maximization (EM) algorithm with convergence properties  [35-36]. Although the multi-links and self-links would be generated because of Poisson distribution, computations become easier without affecting the fundamental outcome significantly [31]. This is due to the influence of calculation error about multi-links and self-links will vanish as the size $n$ of the sparse network becomes large [31]. For using a Poisson distribution on node attributes, we assume that node attributes are high dimensional and sparse [37]. This is common in a lot of real networks such as paper citation or coauthor networks, user relationship networks on social platforms, etc. In addition, the proposed generative model for links can generate a very wide range of network structures due to block structure assumption. By sharing latent variables for links and node attributes, a probabilistic matrix measuring the relationships of network structures and node attributes is introduced to the joint model, which simply assumes the nodes in the same community share similar node attributes, therefore, is able to capture multiple semantics for a community by analyzing its related attributes. 

The remaining of this paper is organized as follows. In Section 2, we introduce our model. In Section 3, the model parameters are estimated by expectation-maximization (EM) algorithm. In Section 4, we evaluate the newly proposed model on a number of artificial benchmarks and real networks with various structures. In Section 5, we draw our conclusions.

\section{A Generative Model on Attributed Networks}
In this section, a joint  generative model will be proposed for undirected networks with node attributes. Firstly, we will introduce the process for generating links based on the block structure assumption [32] and the idea of link communities [33-34]. Then, the hypothesis and the process for generating node attributes will be described.

Given an attributed network $G(V, E, W)$ with $n$ nodes $V=\{ 1, 2, \dots, n \}$, $m$ links $E=\{e_1, e_2, \dots, e_m\}$ and $K$ node attributes, the network is usually represented by an adjacent matrix $A=(A_{ij})_{n \times n}$ and an attribute matrix $W=(W_{ik})_{n \times K}$, where $A_{ij}=1$ if a link exists between nodes $i$ and $j$,  or 0 otherwise; $W_{ik}=1$ if node $i$ has the $k$-th attribute, or 0 otherwise. Suppose the network $A$ has $c$ distinct link communities. The nodes connected to a link community (i.e., a set of closely interrelated links [33]) form a collection of nodes which we call a deduced node community $V_r \;(r = 1,2, \cdots ,c)$, then we have $ V = \bigcup\limits_{r = 1}^c {V_r }$ and each node can belongs to multiple node communities on account of link communities [33, 38].

In a standard stochastic blockmodel [32], a probability matrix $ \Theta = ( \theta _{rs})_{c \times c}$ controls the probabilities for generating links in a network, where $\theta _{rs}$ is the connecting probability of two nodes $i\in V_r$ and $j\in V_s$ and is apparently only related to the communities to which $i$ and $j$ belong. In this study, this constraint is relaxed by introducing a parameter matrix $ D = \left( {d_{ir} } \right)_{n \times c} $, where $ d_{ir}$ is the probability that a node $i$ belongs to the $r$-th deduced community $V_r$ and $ \sum\limits_{i = 1}^n {d_{ir} } = 1 $. We then use $D$ and $\Theta$ together to generate the expected adjacency matrix $ \hat{A} $ of a network $ G $. Specifically, $ \hat A_{ij}^{rs}  = d_{ir} \theta _{rs} d_{js} $ is the expected number of links that lie between nodes $i$ and $j$ in communities $V_r$ and $V_s$, respectively. Summing over communities $V_r$ and $V_s$, the expected total number of links between nodes $i$ and $j$ is $ \hat A_{ij}  = \sum\limits_{rs} {d_{ir} \theta _{rs} d_{js} } $. This generative process is similar with the model in [30]. Suppose the generation of links is independent of each other and the real number of links follows a Poisson distribution with mean value $ \hat A_{ij} $, we have
$$
\begin{array}{l}
\displaystyle \Pr \left( {A|D,\Theta } \right) = \prod\limits_{i < j} {\frac{{\left( {\displaystyle \sum\limits_{r,s} {d_{ir} \theta _{rs} d_{js} } } \right)^{a_{ij} } }}{{a_{ij} !}}\exp \left\{ { - \sum\limits_{r,s} {d_{ir} \theta _{rs} d_{js} } } \right\}}  \\ \vspace{1mm}
\displaystyle \;\;\;\;\;\;\;\;\;\;\;\;\;\;\;\;\;\;\;\;\;\;\; \times \prod\limits_i {\frac{{\left( {\displaystyle \frac{1}{2}\sum\limits_r {d_{ir} \theta _{rr} d_{ir} } } \right)^{a_{ii} /2} }}{{\left( {a_{ii} /2} \right)!}}\exp \left\{ { - \frac{1}{2}\sum\limits_r {d_{ir} \theta _{rr} d_{ir} } } \right\}} . \\ 
\end{array}  \eqno(1)
$$

This model inherits an advantage of the standard stochastic blockmodel that can produce a wide variety of network structures. For example, small off-diagonal elements and big diagonal elements of $\Theta$ would generate traditional community structure (i.e., a set of communities with dense internal connections and sparse external ones). Other choices of probability matrix can generate multipartite, hierarchical, or core-periphery structures, etc.

Now, it's time for the generative process of node attributes. Usually, the attributes of nodes in a network are richness so that the dimensionality of attributes is high, but few nodes can have so many attributes at the same time. In other words, the attributes of nodes are somewhat sparse. Generally, suppose $W_{ik} (k=1,2,\dots,K)$ is independent and identically distributed and is a binomial distribution ($W_{ik}=0$ or $1$), high dimensionality and sparsity of node attributes mean that the dimension $K$ is large and the probability of $W_{ik}=1$ in the binomial distribution is small. Thus, by Poisson limit theorem [37], a Poisson distribution can be used to generate node attributes. It is believed that nodes in the same community share similar attributes. Let $ \phi _{rk} $ denote the probability that community $V_r$ has the $k$-th attribute, $ \sum\limits_{k = 1}^K {\phi _{rk} }  = 1$ and $ \Phi  = \left( {\phi _{rk} } \right)_{c \times K} $, then the propensity of a node $i$ in community $V_r$ possessing $k$-th attribute can be represented as $ \hat W_{ik}^r  = d_{ir} \phi _{rk} $. Summing over all communities $V_r$, the mean propensity of a node $i$ possessing $k$-th attribute is $ \hat W_{ik}  = \sum\limits_r {d_{ir} \phi _{rk} } $, and we have 
$$
W_{ik}  \sim {\text{Poisson}}(\hat W_{ik} ) = {\text{Poisson}}\left( {\sum\limits_r {d_{ir} \phi _{rk} } } \right). \eqno(2)
$$

By sharing latent variables for links and node attributes, the generative model for generating topology structure and node attributes can be described as follows:
$$
\begin{array}{l}
\displaystyle \Pr \left( {A,W|D, \Theta ,\Phi } \right) = \prod\limits_{i < j} {\frac{{\left( {\displaystyle \sum\limits_{r,s} {d_{ir} \theta _{rs} d_{js} } } \right)^{a_{ij} } }}{{a_{ij} !}}\exp \left\{ { - \sum\limits_{r,s} {d_{ir} \theta _{rs} d_{js} } } \right\}}  \\ 
\displaystyle \;\;\;\;\;\;\;\;\;\;\;\;\;\;\;\;\;\;\;\;\;\;\;\;\;\;\;\;\;\;\;\;\;\; \times \prod\limits_i {\frac{{\left( {\displaystyle \frac{1}{2}\sum\limits_r {d_{ir} \theta _{rr} d_{ir} } } \right)^{a_{ii} /2} }}{{\left( {a_{ii} /2} \right)!}}\exp \left\{ { - \frac{1}{2}\sum\limits_r {d_{ir} \theta _{rr} d_{ir} } } \right\}}  \\ 
\displaystyle \;\;\;\;\;\;\;\;\;\;\;\;\;\;\;\;\;\;\;\;\;\;\;\;\;\;\;\;\;\;\;\;\;\; \times \prod\limits_{i,k} {\frac{{\left( {\displaystyle \sum\limits_r {d_{ir} \phi _{rk} } } \right)^{W_{ik} } }}{{W_{ik} !}} \exp \left\{ - \sum\limits_r {d_{ir} \phi _{rk} } \right\} }, \\ 
\end{array} \eqno(3)
$$
where $ \Theta $ is symmetrical and $ \sum\limits_{r,s} { \theta _{rs} }=1 $. Unlike PPSB\_DC [16] the attribute model in Eq.(3) still follows Poisson generative process where the parameter $\Phi$ quantifies how much the network structures depend on node attributes. Meanwhile, the attributes that are closely related to a community naturally represent the semantic of the community. Such semantics can help explain why certain nodes belong to a community. And the semantics are complementary to structural information for forming network structures.

\section{EM Algorithm for Inferring the Model Parameters}
The model in Eq.(3) is specified by three types of parameters. The first is the observed data: the adjacent matrix $A$ and the attribute matrix $W$. The second is the latent data: the cluster label of nodes which takes value $1, 2, \dots, c$. The third is the model parameters: $D$, $\Theta$ and $\Phi$. Neglecting constants and terms independent of the parameters, the logarithm of Eq.(3) can be expressed as follows: 
$$
\begin{array}{l}
\displaystyle L\left( {D,\Phi ,\Theta } \right) = \sum\limits_{i,j} {\left[ {\frac{1}{2}a_{ij} \log \left( {\sum\limits_{r,s} {d_{ir} \theta _{rs} d_{js} } } \right) - \frac{1}{2}\sum\limits_{r,s} {d_{ir} \theta _{rs} d_{js} } } \right]}  \\ 
\vspace{2mm}
\;\;\;\;\;\;\;\;\;\;\;\;\;\;\;\;\;\;\;\;\;\;\; + \displaystyle \sum\limits_{i,k} {\left[ {W_{ik} \log \left( {\sum\limits_r {d_{ir} \phi _{rk} } } \right) - \sum\limits_r {d_{ir} \phi _{rk} } } \right]} . \\ 
\end{array} \eqno(4)
$$

Our goal is to infer group memberships of nodes, i.e., the probabilities of nodes belong to community $V_r$ ($r=\{1,2,\cdots,c\}$). Unfortunately, we cannot measure them directly because they are hidden or latent data. In this study, an expectation-maximization (EM) algorithm [35] that is convergent and can easily handle models with hidden variables is used to optimize the joint log-likelihood function.

In E-step,  given the current parameters $D$, $\Theta$ and $\Phi$, calculating an expected value $\bar{L}$ for the log-likelihood by averaging over latent variables, we have
$$
\begin{array}{l}
\displaystyle \bar L\left( {D,\Phi ,\Theta } \right) = \frac{1}{2}\sum\limits_{ijrs} {\left[ {a_{ij} q_{ij}^{rs} \log \left( {\frac{{d_{ir} \theta _{rs} d_{js} }}{{q_{ij}^{rs} }}} \right) - d_{ir} \theta _{rs} d_{js} } \right]}  \\ 
\;\;\;\;\;\;\;\;\;\;\;\;\;\;\;\;\;\;\;\;\; + \displaystyle \sum\limits_{ikr} {\left[ {W_{ik} \gamma _{ik}^r \log \left( {\frac{{d_{ir} \phi _{rk} }}{{\gamma _{ik}^r }}} \right) - d_{ir} \phi _{rk} } \right]} , \\ 
\end{array} \eqno(5)
$$
where
$$
q_{ij}^{rs}  = \frac{{d_{ir} \theta _{rs} d_{js} }}{{ \sum\limits_{rs} {d_{ir} \theta _{rs} d_{js} } }},\;\gamma _{ik}^r  = \frac{{d_{ir} \phi _{rk} }}{{ \sum\limits_r {d_{ir} \phi _{rk} } }} \eqno(6)
$$
are the expected probabilities of between nodes $i ( \in V_r ) $ and $j ( \in V_s ) $ to be linked and those of $i ( \in V_r ) $ possessing $k$-th attribute, respectively. In fact, $\bar{L}(D, \Theta, \Phi)$ is the lower bound of $L(D, \Theta, \Phi)$ according to the Jensen's inequality.

In M-step, given values of $ q_{ij}^{rs}$ and $\gamma _{ik}^r$, we can obtain the estimates of $D$, $\Theta$ and $\Phi$ that maximize the expected log-likelihood $\bar{L}(D, \Theta, \Phi)$ in Eq.(5) by using Lagrange multiplicator method in the following. 
$$
d_{ir}  = \frac{{ \sum\limits_{js} {a_{ij} q_{ij}^{rs} }  + \sum\limits_k {W_{ik} \gamma _{ik}^r } }}{{\sum\limits_{ijs} {a_{ij} q_{ij}^{rs} }  + \sum\limits_{ik} {W_{ik} \gamma _{ik}^r } }},\;\theta _{rs}  = \frac{{\sum\limits_{ij} {a_{ij} q_{ij}^{rs} } }}{{\sum\limits_{ijrs} {a_{ij} q_{ij}^{rs} } }},\;\phi _{rk}  = \frac{{\sum\limits_i {W_{ik} \gamma _{ik}^r } }}{{\sum\limits_{ik} {W_{ik} \gamma _{ik}^r } }}. \eqno(7)
$$

\begin{algorithm}[H]
	\caption{EM algorithm for PSB\_PG}
	\label{Al:01}
	\renewcommand{\algorithmicrequire}{ \textbf{Input:}} 
	\renewcommand{\algorithmicensure}{ \textbf{Output:}} 
	\begin{algorithmic}[1]
		\REQUIRE ~~ \\
		the adjacency matrix $A$ \\
		the attribute matrix $W$ \\
		the number of communities $c$ \\
		the maximum iteration $I_{T}$ and the threshold $\epsilon$
		\ENSURE ~~\\
		the inferred parameters $D, \Theta, \Phi$
		\STATE Initialize  $D^{(0)}, \Theta^{(0)}, \Phi^{(0)}$
		\STATE Compute objective function $	L^{(0)} \left( {D^{(0)} ,\Phi ^{(0)} ,\Theta ^{(0)} } \right) $ by Eq.(4)
		\FOR{$t=1:I_{T}$}
		\vspace{2mm}
		\STATE E-step: Compute $ q_{ij}^{rs} ,\;\gamma _{ik}^r  $ by Eq.(6)
		\vspace{1mm}
		\STATE M-step: Compute $ D^{(t)} ,\;\Theta ^{(t)} ,\;\Phi ^{(t)} $ by Eq.(7)
		\vspace{1mm}
		\STATE Compute objective function $	L^{(t)} \left( {D^{(t)} ,\Phi ^{(t)} ,\Theta ^{(t)} } \right) $ by Eq.(4)
		\vspace{1mm}
		\IF{ $ \left| {L^{(t)} \left( {D^{(t)} ,\Phi ^{(t)} ,\Theta ^{(t)} } \right) - L^{(t - 1)} \left( {D^{(t - 1)} ,\Phi ^{(t - 1)} ,\Theta ^{(t - 1)} } \right)} \right| < \epsilon$ or $t=I_{T}$}
		\vspace{1mm}
		\STATE $ D = D^{(t)} ,\;\Theta  = \Theta ^{(t)} ,\;\Phi  = \Phi ^{(t)} $; Break;
		\ENDIF
		\ENDFOR
	\end{algorithmic}
\end{algorithm}

The derivation of Eq.(7) can be found in Appendix A. The right term in $d_{ir}$ of Eq.(7) visually shows how the node attributes helps to enforce the intra-cluster similarity. Iteratively updating Eq.(6) and Eq.(7) guarantees to find a local optimum of the lower bound $\bar{L}(D, \Theta, \Phi)$ [36]. The detail process for inferring the model parameters is showed in Algorithm 1. For simplicity, we call the proposed model PSB\_PG (\underline{P}oisson general \underline{s}tochastic \underline{b}lock model for links coupling \underline{P}oisson \underline{g}enerative model for attributes).\\

\noindent {\bf Initialized scheme of $ \Theta $.} In the algorithm of PSB\_PG, the initial values of the matrix $ \Theta $ will strongly affect the convergence speed of the algorithm. We know that $\Theta$ reflects network structure contained in the network. When the initial values of $ \Theta $ are consistent with the actual network structure, the algorithm will converge quickly. When the initial network structure (i.e., the initial values of $\Theta$) violates the actual network structure, the algorithm will converge slowly and might reach to the local optimal. Therefore, we use maximum entropy distribution [39] and the idea of maximum likelihood to design the initialized scheme of $\Theta$. In detail, given random values of parameters with the exception of $\Theta$, several runs with small number of iterations are performed in three schemes of $\Theta$: the diagonal elements are more large, the off-diagonal elements are more large, and all elements are almost equal. Then, the average of likelihood is calculated for each scheme. The scheme with the largest average likelihood is used to initialize $ \Theta $ in our algorithm.

The time complexity of PSB\_PG algorithm for fitting of our model is mainly controlled by E-step and M-step. For each iteration in this algorithm, the time complexity to evaluate E-step is $ O(mc^2  + nKc) $, where $m$ is the number of links, the time complexity to evaluate M-step is $ O\left( {nc + c^2  + nKc} \right) $. As the number of communities is much smaller than the number of nodes, i.e., $c \ll n$, the time complexity of M-step can be written as $ O( nKc) $. Then, the total time complexity of the algorithm is $ O\left( {I_T \left( {mc^2  + nKc} \right)} \right) $.

\section{Experiments}
In this section, we will first evaluate the performance of our model PSB\_PG on synthetic networks. Through a case study, we then assess whether the parameter $ \Phi  = \left( {\phi _{rk} } \right)_{c \times K} $ can capture the relationships between communities and node attributes. Finally, we will compare the proposed model PSB\_PG with four related methods mentioned in Introduction section: PPSB\_DC [16], BNPA [12], NEMBP [28] and PCL\_DC [20] on artifical and real networks with various structures. Where, by integrating node attributes into the models, PPSB\_DC, BNPA and NEMBP have the ability to detect general structures, PCL\_DC is good at identifying classical community structures.  

For networks with disjoint communities, the widely used Normalized Mutual Information (NMI) [40] index is adopted to measure the accuracy of each method. For networks with overlapping communities, the generalized NMI (GNMI) [41-42] is used. The two accuracy metrics are defined as follows.
$$
NMI(V^{(T)} ,V^{(I)} ) = \frac{H(V^{(T)})+H(V^{(I)})-H(V^{(T)}, V^{(I)})}{(H(V^{(T)})+H(V^{(I)}))/2}, \eqno(8)
$$
$$
GNMI\left( {V^{\left( T \right)} ,\;V^{\left( I \right)} } \right) = \frac{{\displaystyle \frac{1}{2}\left[ {H\left( {V^{\left( T \right)} } \right) - H\left( {V^{\left( T \right)} |V^{\left( I \right)} } \right) + H\left( {V^{\left( I \right)} } \right) - H\left( {V^{\left( I \right)} |V^{\left( T \right)} } \right)} \right]}}{{\max \left\{ {H\left( {V^{\left( T \right)} } \right),\;H\left( {V^{\left( I \right)} } \right)} \right\}}}. \eqno(9)
$$
Where $V^{(T)} = \left( {V_{1}^{(T)} ,V_{2}^{(T)} , \cdots ,V_{c}^{(T)} } \right)$ and $V^{(I)} = \left( {V_{1}^{(I)} ,V_{2}^{(I)} , \cdots ,V_{c}^{(I)} } \right)$  are true communities and inferred communities given a network, respectively; $ H(V^{(T)}) (H(V^{(I)} )) $ is the entropy of the partition $V^{(T)} (V^{(I)} ) $; $ H(V^{(T)}, V^{(I)}) $ is the joint entropy; $H(V^{(T)} \mid V^{(I)})$ is the conditional entropy inferring $V^{(T)}  $ given $V^{(I)} $ and vice versa. A larger NMI means a better partition for a disjoint partition, and a larger GNMI means a better partition for a overlapping partition.

\subsection{Effectiveness of PSB\_PG on Synthetic Networks}
In this section, the efficiency of the proposed method is tested on artificial benchmarks with various structures including non-overlapping community structures, overlapping community structures, bipartite structures, community structures with multiple attribute semantics. Since generative models are sensitive to their initial values, we run our model 30 times and report the average results.

{\bf LFR benchmarks with disjoint communities.} Following the parameters used in [43], we generated networks using the following parameter settings: $\{N=500, k=15, maxk=45, \mu=0.1 \sim 0.9, minc=20, maxc=50\}$, 
where $N$ is the number of nodes, $k$ is the average degree of nodes, $maxk$ is the maximum degree of nodes, $\mu$ is the mixing parameter, $minc$ is the minimum for the community sizes, $maxc$ is the maximum for the community sizes. The strength of network structure is controlled by mixing parameter $\mu$, which is the fraction of nodes connect to nodes in the other communities. The smaller $\mu$, the clearer the community structure is. The distributions of degrees and community sizes are power laws with exponents $\eta=2$ and $\beta=1$, respectively. Under this parameter setting, we generate a batch of artificial networks with $\mu=0.1, 0.2,\cdots,0.9$ and each network contains 14 communities (i.e., $c=14$).

Given a LFR network, we then use the following strategy to generate node attributes. Assuming that each community has strong correlation with $h=5$ binary attributes (the correlation is controlled by the probability $p_{in}$) and weak correlation with the rest $(c-1)h=(14-1)h=65$ binary attributes (this correlation is controlled by the probability $p_{out}$). Thus, each node in community $V_r$ has $70$ attributes. Given a community $V_r$, $h =5$ attributes with strong correlation were generated by $ W_{i,5}^r  \sim Ber(5,p_{in} ) $, the remaining $65$ attributes with weak correlation were generated by $ W_{i,65}^{s (s \neq r)}  \sim Ber(65,p_{out} ) $. Fixed $ p_{out}  = 0.1$, let $ p_{in} $ range from 0.6 to 0.9. We generated 4 groups of attributed networks. The larger $p_{in}$, the tighter relationships between communities and node  attribute are. 

\begin{figure}[ht]
	\centering
	\includegraphics[width=8.5cm]{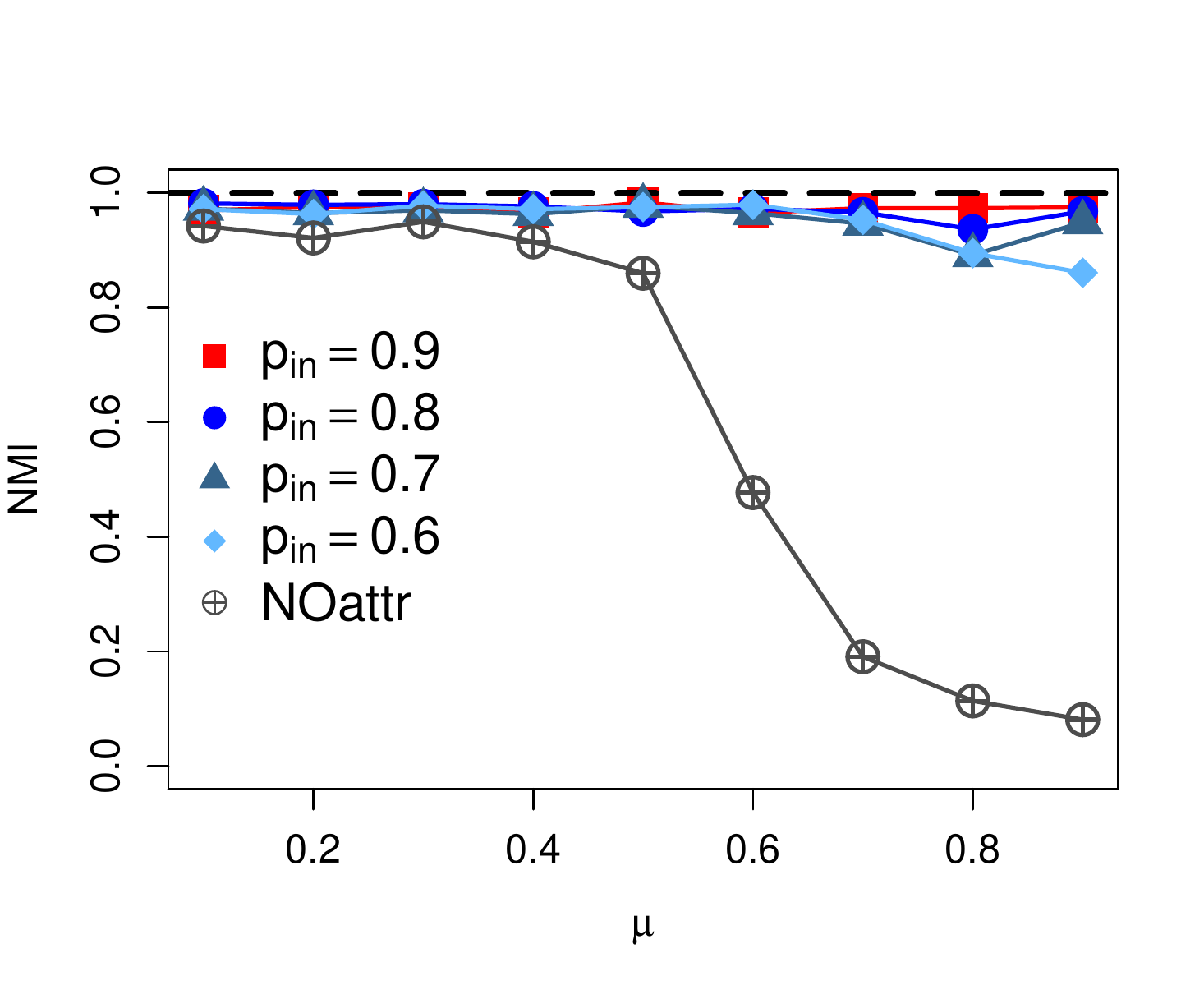}
	\caption{The performance of the proposed method PSB\_PG on LFR networks with disjoint communities. The average NMI for each $\mu$ was calculated from 30 runs. "NOattr" represents the identification results on networks with no attributes.}\label{fig1}
\end{figure}

On these attributed networks, the community identification results of PSB\_PG were shown in Figure \ref{fig1}.
As can be seen from Figure \ref{fig1}, overall, integrating links of a network and node attributes will significantly boost the performance of community detection, especially when $\mu \geq 0.6$. This can be easily explained by the right term of $d_{ir}$ in Eq.(7). Namely, when network structure is vague, the attribute information $ \sum\limits_k {W_{ik} \gamma _{ik}^r } $ is very useful to improve the accuracy of node assignments.

{\bf LFR benchmark with overlapping communities.} It is well known LFR generator can generate overlapping communities. In this study, we set $om=2$ (which is the number of memberships of the overlapping nodes) and $on=50, 100$, or 150 (which is the number of overlapping nodes), therefore, the fraction of overlapping nodes is $10\%$, $20\%$ or $30\%$. The rest parameters are the same as the ones of LFR benchmark with non-overlapping communities in the previous experiments. 

\begin{figure}[ht]
	\centering
	\includegraphics[width=16.5cm]{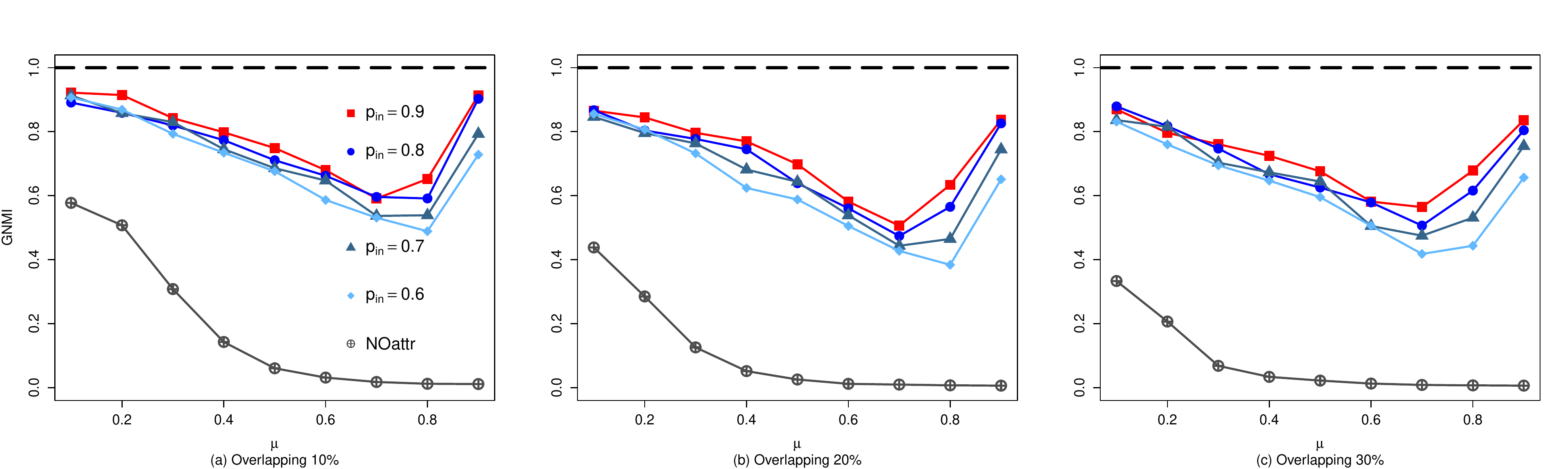}
	\caption{The performance of the proposed method PSB\_PG on LFR networks with overlapping communities. The average GNMI for each $\mu$ was calculated from 30 runs. "NOattr" represents the identification results on networks with no attributes.}\label{fig2}
\end{figure}

On these attributed networks, the overlapping community identification results of PSB\_PG were shown in Figure \ref{fig2}, where we use the diagonal element of $ \Theta  = \left( {\theta _{rs} } \right)_{c \times c} $ as a threshold, if $d_{ir}>\theta_{rr}$, we assigned the node $i$ to the community $V_r$.

As can be seen from Figure \ref{fig2}, the identification results on networks with node attributes are better than the ones without attributes. The bigger $p_{in}$, the better accuracy is. For the mixing coefficient $ \mu $=0.8 or 0.9, the network structures are not easy to be identified by any existing overlapping community detection method using topology merely, but our proposed method works well when taking node attributes into account.

\begin{figure}[ht]
	\centering
	\includegraphics[width=14.5cm]{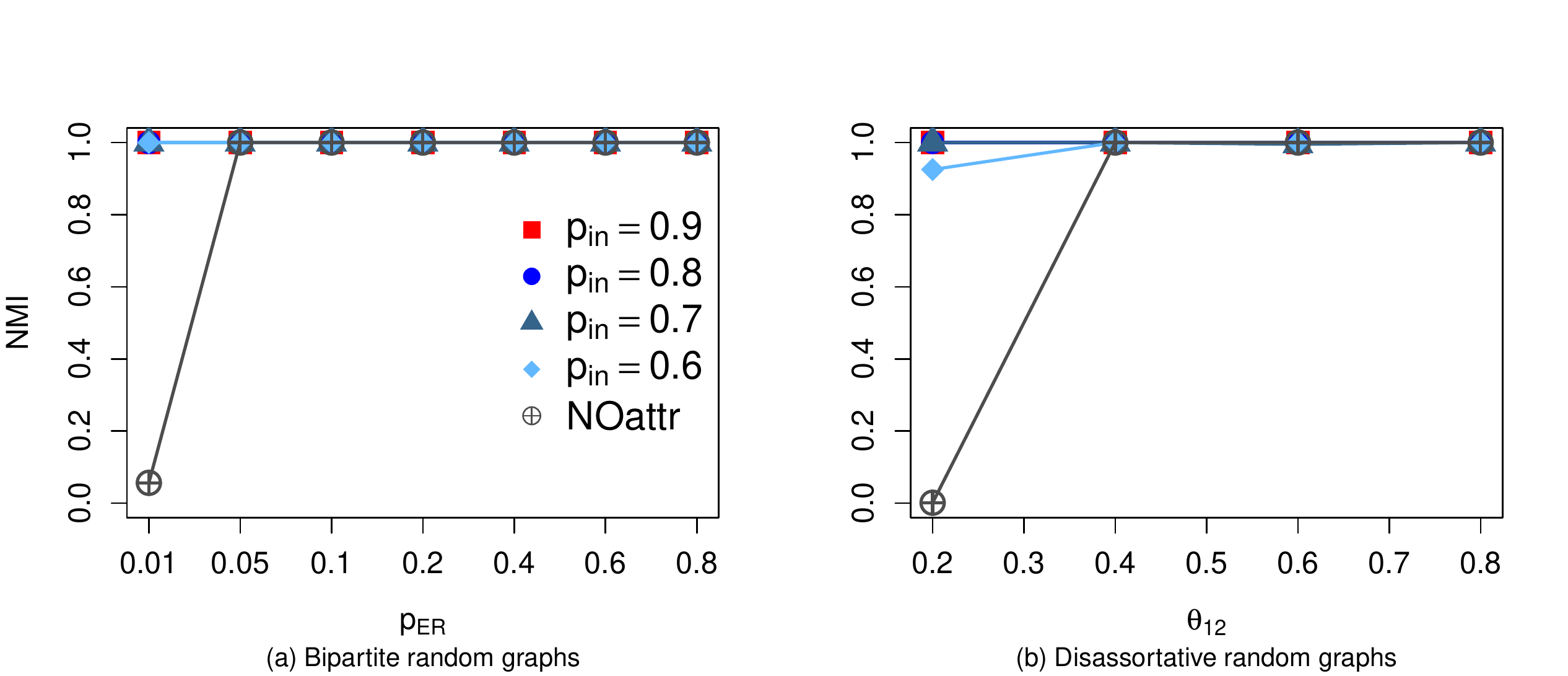}
	\caption{The performance of the proposed method PSB\_PG on (a) bipartite networks, (b) disassortative networks. The average NMI of each point was calculated from 30 runs. "NOattr" represents the calculated results with no attributes.}\label{fig3}
\end{figure}

{\bf Bipartite networks.} In this group of experiments, we evaluated the performance of the proposed method on other network structures. We used Bipartite networks as an example, 
which were generated by ER (Erd$\ddot{o}$s-R$\acute{e}$nyi) model [44]. In detail, each network has two subgraphs with size 200 and 300, respectively. There are no links within subgraph. And the connecting probability between-subgraphs is $p_{ER}  = 0.8,0.6,0.4, 0.2,0.1,0.05$ or $0.01$.
Similarly, we assume each subgraph has strong correlation with $h =5$ attributes at the strength $p_{in}=0.6,\dots, 0.9$ and has weak correlation with the other. The bipartite identification results of PSB\_PG on these attributed networks were shown in Figure \ref{fig3}.a.

From Figure \ref{fig3}.a, the proposed method PSB\_PG works very well on the bipartite networks with or without attribute information with the exception of $p_{ER}  = 0.01$ under the condition no node attributes attached to the networks. At this situation, the bipartite structure is too vague to be detected. However, with the help of node attributes, the structure can be easily detected. 

In Figure \ref{fig3}.b, we further showed the results of our method when a small number of links were generated within each subgraph of a bipartite structure. These networks (i.e., disassortative networks [13]) in Figure \ref{fig3} (b) were generated using the idea of stochastic blockmodel, and the probability matrix of links between communities is
$$
\Theta  = \left( {\theta _{rs} } \right)_{c \times c}  = \left( {\theta _{rs} } \right)_{2 \times 2}  = \left( \begin{array}{l}
0.2\;\;\;\;\theta _{12}  \\ 
\theta _{12} \;\;\;0.2 \\ 
\end{array} \right),\eqno(10)
$$
where $ \theta_{12}=0.8,0.6,0.4$ or $0.2$. The size of networks are still 500, and two subgraphs are also 200 and 300, respectively.

As it is shown in Figure \ref{fig3}.b, when $ \theta_{12}=0.2$, network structures are too vague to be detected, but at the help of node attributes, the network structures still can be almost exactly detected by our model. 

\begin{figure}[ht]
	\centering
	\begin{minipage}[c]{0.8\textwidth}
		\centering
		\includegraphics[width=12.5cm]{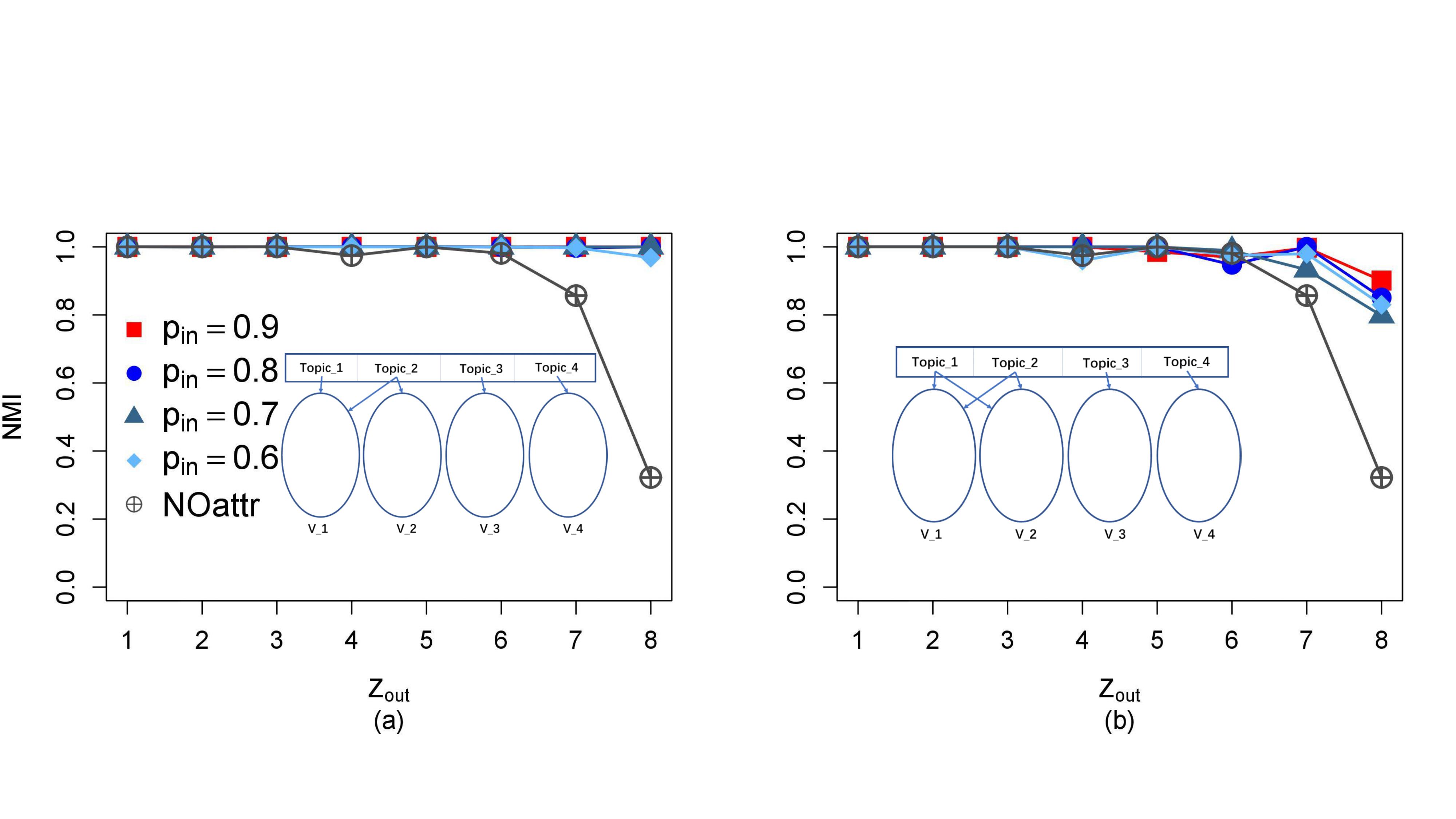}
	\end{minipage}%
	
	\begin{minipage}[c]{0.8\textwidth}
		\centering
		\includegraphics[width=12.5cm]{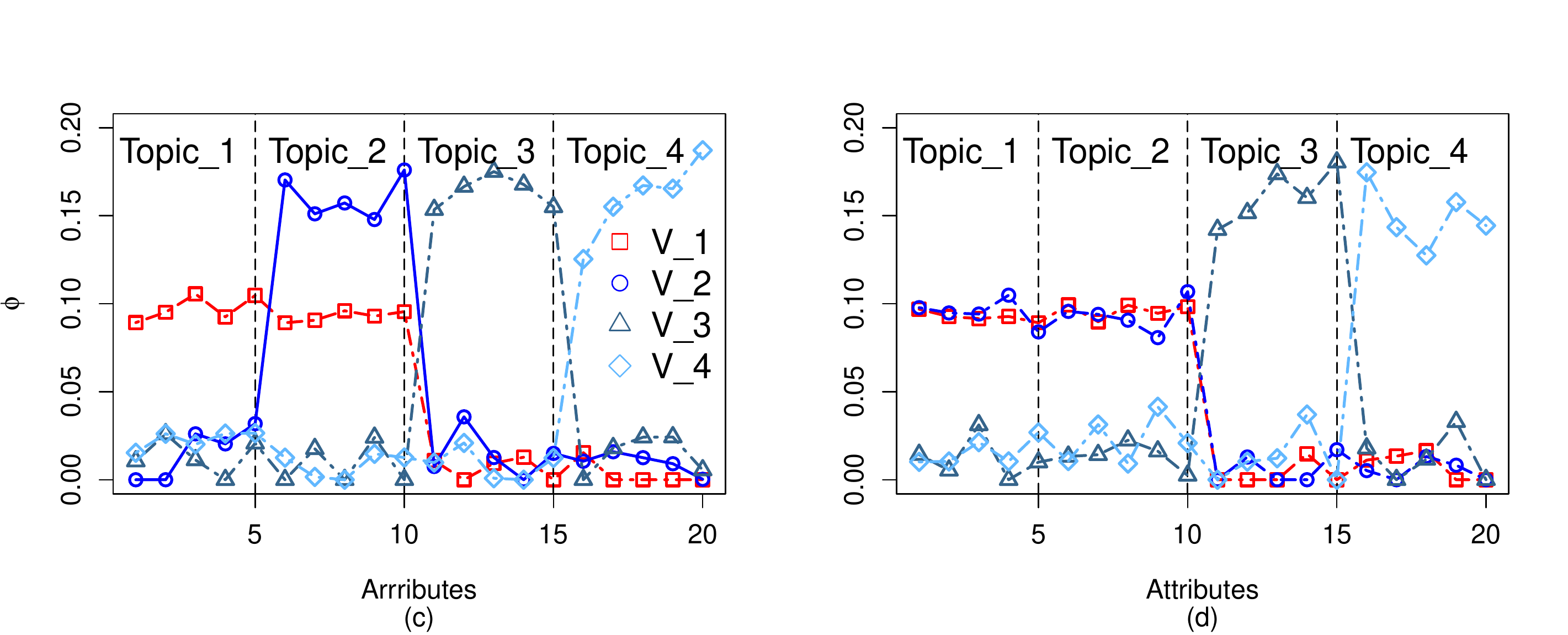}
	\end{minipage}
	\caption{The performance of the proposed model PSB\_PG on GN networks when the memberships of communities and attributes are inconsistent. "Topic\_1", "Topic\_2","Topic\_3" and "Topic\_4" stand for semantic topics. Each topic contains five attributes. "V\_1", "V\_2", "V\_3" and "V\_4" represents communities. For (a) and (b), the average NMI was calculated from 30 runs. "NOattr" represents the experimental results on networks without node attributes. For (c) and (d), each element of $ \Phi  = \left( {\phi _{rk} } \right)_{c \times K} $  quantifies how much the community V\_r relies on the $k$-th attribute.}\label{fig4}
\end{figure}

{\bf GN networks with multiple semantics for each community.} In this group of experiments, we used GN-type networks [2] as the basis, which can also be generated by LFR generator. Where each network contains 4 communities. The generated network structure, their related topics and the identification results were showed in Figure \ref{fig4}.a-d. 

In Figure \ref{fig4}.a, we assume community V\_1 have 2 topics (i.e., Topic\_1 and Topic\_2), and each of the other three communities V\_2, V\_3 and V\_4 is related to the topic Topic\_2, Topic\_3 or Topic\_4, respectively. In Figure \ref{fig4}.b, we assume communities V\_1 and V\_2 share the same 2 topics: Topic\_1 and Topic\_2, the other two  V\_3 or V\_4 only contain one disjoint topic Topic\_3 or Topic\_4. Each topic contains 5 attributes showed in Figure \ref{fig4}.c-d. As shown in Figure \ref{fig4}.a-b, adding node attributes promotes the performance of community detection, especially when network structure is not clear ($Z_{out} > 6$). The results showed in Figure \ref{fig4}.a-b have demonstrated that the inconsistent memberships of communities and attributes has little effect on the community detection ability of PSB\_PG. 

In Figure \ref{fig4}.c-d, $ \Phi  = ( \phi _{rk})_{c \times K} $ corresponds to the block matrix of Figure \ref{fig4}.a-b  ($ z_{out}=1, p_{in}=0.9$), respectively, where $\phi _{rk}$ quantifies how much the community V\_r depends on the $k$-th attribute. The stronger the relevance of communities and topics, the greater the corresponding value of $ \phi $. For example, in Figure \ref{fig4}.c, the inferred $\phi _{1k} (k=1,2,\dots,10)$ is significantly larger than the other $\phi _{1k} (k=11,12,\dots,20)$ because, the community V\_1 is strongly related to Topic\_1 (including 1-5 attributes) and Topic\_2 (including 6-10 attributes), and weakly related to Topic\_3 and Topic\_4. Similarly, in Figure \ref{fig4}.d, the inferred $\phi _{1k} (k=1,\dots,10)$ and $\phi _{2k} (k=1,\dots,10)$ are significantly larger than the other $\phi _{1k} (k=11,12,\dots,20)$ and $\phi _{2k} (k=11,12,\dots,20)$ because community V\_1 and community V\_2 are strongly related to Topic\_1 and Topic\_2. This phenomenon has illustrated that the newly proposed model PSB\_PG is able to capture the relationships between communities and attributes, whenever they are consistent or inconsistent.

\subsection{The Interpretability of Inferred Communities: A Case Study} 

\begin{figure}[h]
	\centering
	\includegraphics[width=12.5cm]{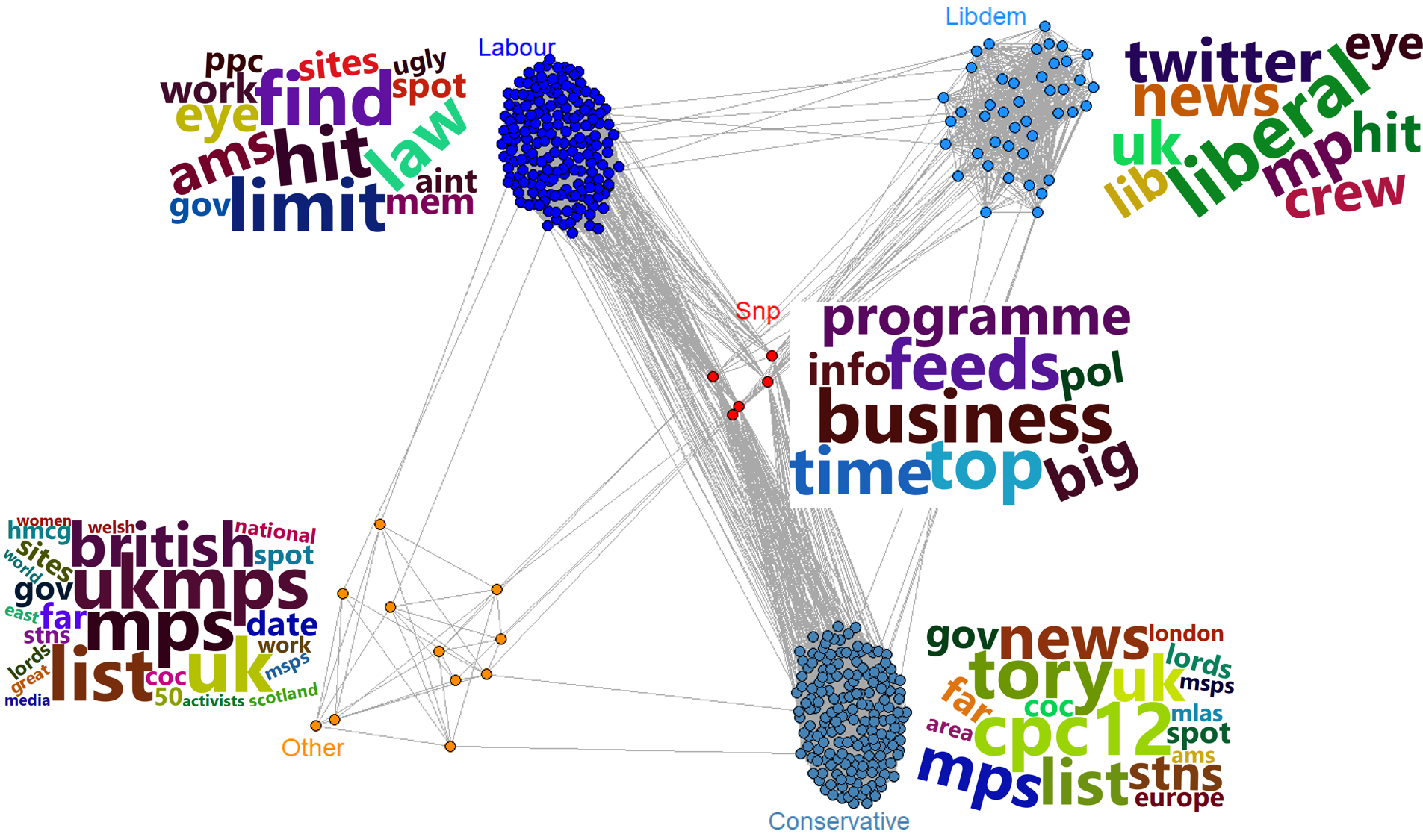}
	\caption{The inferred semantic interpretation of communities on the "politics-uk" dataset. Users are coloured and labelled based on the ground truth which consists of five groups, corresponding to five different political parties. The inferred top 100 attributes are placed beside each community.}\label{fig5}
\end{figure}

In this section, by a case study, we intend to show the interpretability of the newly proposed model PSB\_PG by analyzing the model parameter $\Phi  = \left( {\phi _{rk} } \right)_{c \times K}$ on the real Twitter dataset: "politics-uk" [45]. The "politics-uk" is a collection of 419 users, corresponding to 419 members of Parliament in the United Kingdom. The ground truth consists of five groups, corresponding to political parties: Conservative, Labour, Liberal Democrat (Libdem), Scottish National Party (Snp) and other (see Figure \ref{fig5}). The 419 active users on Twitter post 539,592 tweets and contain 3,614 Twitter lists. The links are constructed by the users whom they follow. The attribute information of each user covers a vector constructed from the aggregation of both "names" and "descriptions" of the 500 Twitter lists to which each user has most recently been assigned. The dimension of attributes of each user is 2879. The data files can be downloaded from http://mlg.ucd.ie/aggregation/index.html.

We select the top 100 attributes for each community according to their values of inferred $ { \phi _{rk} } $. The semantics implied in each community can be visualized by Figure \ref{fig5}. We then have the semantic interpretation of each community by using its mostly related attributes. Taking political party Labour as an example, the users in this party are more concerned about working and government strategies. 
In general, the parameter matrix $ \Phi  = \left( {\phi _{rk} } \right)_{c \times K} $ of our model is capable of capturing the relationships between communities and attributes. These inferred attributes can help us to understand the semantics of each community.

\subsection{The Comparison of models on Artificial and Real Networks} 
In this section, we will compare the proposed model with the state-of-the-art generative models PPSB\_DC [16] BNPA [12], NEMBP [28], PCL\_DC [20] on artificial networks and real networks with various structures.

First, we have compared these models on artificial networks contained different types of structures including LFR6\_community, LFR7\_community, LFR8\_community, ER\_Bipartite, SBM\_Disassortative, SBM\_Mixture GNmulti-semantics. LFR6\_community, LFR7\_community and LFRm8\_community stand for LFR networks with community structures in Figure \ref{fig1} when $ \mu =0.6, 0.7, 0.8$ and $ p_{in}=0.6$. ER\_Bipartite and SBM\_Disassortative represent the networks with bipartite structure and disassortative structure corresponding to Figure \ref{fig3}.a-b ($ p_{in}=0.6$). SBM\_Mixture stands for a network with mixture structure generated by SBM with block matrix
\[
\Theta  = \left( \begin{array}{l}
0.10\;\;0.40\;\;0.10 \\ 
0.40\;\;0.05\;\;0.02 \\ 
0.10\;\;0.02\;\;0.01 \\ 
\end{array} \right),
\]
whose sizes of three communities are 80, 100 and 120 respectively, and each community has strong correlation with 5 binary attributes ($ p_{in}=0.6$) and weak correlation with the rest 10 attributes ($ p_{out}=0.1$). GNmulti-semantics stands for the GN network corresponding to Figure \ref{fig4}.b ($ p_{in}=0.9, Z_{out}=1$). 
The average accuracy (measured by NMI) among 30 runs of the compared models are shown in Table \ref{tab0} and the best results are marked in bold.

\begin{table}[htbp]
	\centering
	\caption{Comparison of generative models on synthetic networks}
	\resizebox{\textwidth}{!}{
		\begin{tabular}{lccccc}
			\toprule
			NMI   & PSB\_PG & NEMBP & BNPA  & PPSB\_DC & PCL\_DC \\
			\hline
			(a) LFR6\_community & \textbf{0.9791$ \pm $0.0211} & 0.9394$ \pm $0.0170 & 0.9610$ \pm $0.0022 & 0.8767$ \pm $0.0263 & 0.9779$ \pm $0.0089 \\
			(b) LFR7\_community & \textbf{0.9526$ \pm $0.0236} & 0.9293$ \pm $0.0205 & 0.9319$ \pm $0.0162  & 0.8747$ \pm $0.0498 & 0.8896$ \pm $0.0369 \\
			(c) LFR8\_community & 0.8946$ \pm $0.0279 & 0.9200$ \pm $0.0185 & \textbf{0.9508$ \pm $0.0054} & 0.8667$ \pm $0.0304	& 0.5965$ \pm $0.0422 \\
			\hline
			(d) ER\_Bipartite & \textbf{1.0000$ \pm $0.0000} & \textbf{1.0000$ \pm $0.0000} & \textbf{1.0000$ \pm $0.0000} & 0.9558$ \pm $0.0213 & 0.0325$ \pm $0.0196 \\
			(e) SBM\_Disassortative & \textbf{1.0000$ \pm $0.0000} & \textbf{1.0000$ \pm $0.0000} & 0.0011$ \pm $0.0003 & 0.9600$ \pm $0.0238 & 0.0473$ \pm $0.0528 \\
			(f) SBM\_Mixture & \textbf{0.9481$ \pm $0.0046} & 0.9047$ \pm $0.0000 & 0.6760$ \pm $0.0007   & 0.4579$ \pm $0.0522 & 0.1857$ \pm $0.0489 \\
			\hline
			(g) GNmulti-semantics  & \textbf{1.0000$ \pm $0.0000} & 0.9526$ \pm $0.0857 & 0.8571$ \pm $0.0000 & 0.8409$ \pm $0.0536 & 0.8534$ \pm $0.0136 \\
			\toprule
		\end{tabular}%
	}
	\label{tab0}%
\end{table}%

\begin{figure}[h]
	\centering
	\includegraphics[width=8.5cm]{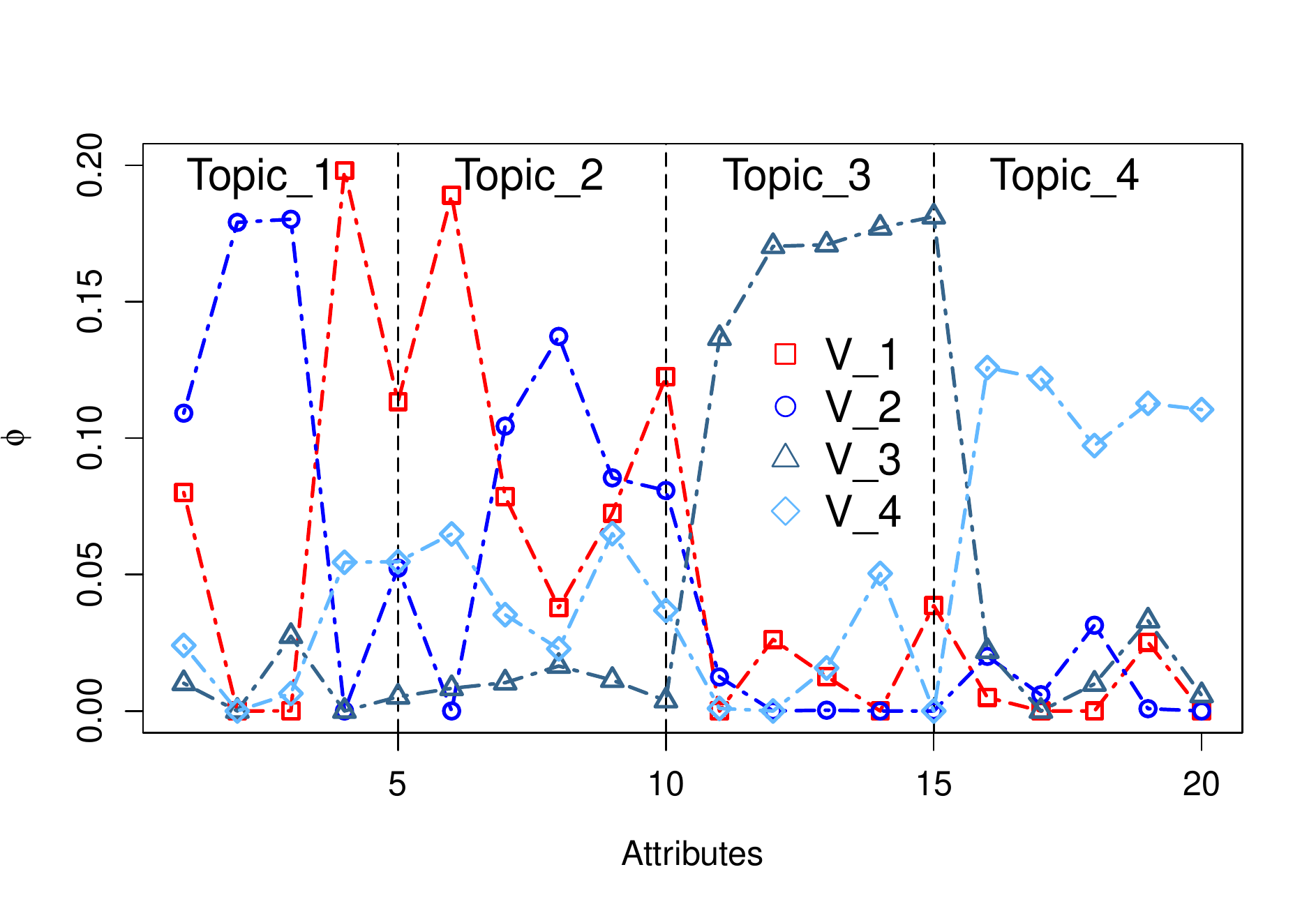}
	\caption{The relationships between communities and attributes inferred by NEMBP. The same relationships inferred by the proposed method PSB\_PG are shown in Figure \ref{fig4}.d }\label{fig7}
\end{figure}

As shown in Table \ref{tab0}, for community structures (a, b, c), all of five models work well with the exception of PCL\_DC on LFR8\_community. Relatively speaking, PSB\_PG, NEMBP and BNPA are better than PPSB\_DC and PCL\_DC. For other network structures (d, e, f), the new model PSB\_PG is superior to other models. Unexpectedly, the performances of BNPA are very poor on networks with mixture structures (i.e., e, f). PCL\_DC shows the worst performance on these networks as expected since it is designed to detect community structures. Moreover, PSB\_PG shows the best performance for detecting structures with multiple semantics. NEMBP shows a good semantic interpretability, but the mixture of topics on communities leads to the worse performance than on communities where each contains a single topic. An illustration example can be shown in Figure \ref{fig7} and Figure \ref{fig4}.d by comparing the predicted relationships between communities and corresponding topics of NEMBP and PSB\_PG. By Figure \ref{fig7} (the communities V\_1 and V\_2 have multiple topics), the relationships between communities and attributes inferred by NEMBP is worse than the ones by the new model PSB\_PG (Figure \ref{fig4}.d).

We then have compared the proposed model with PPSB\_DC, BNPA, NEMBP, PCL\_DC on real networks with mixture structures (the first four networks: Cornell, Texas, Washington, Wisconsin in WebKB datasets (http://www-2.cs.cmu.edu/~webkb/)) and community structures (Cora and Citeseer [46]). The properties of these networks are showed in Table \ref{tab1}, where $n, m$ and  $c$ stand for the number of nodes, links and communities, respectively; $K$ is the dimension of node attributes. On these networks, we run all models 30 times and report their average accuracy (NMI). The experimental results are showed in Table \ref{tab2}. The best results are marked in bold.

\begin{table}[h]
	\centering
	\caption{The properties of real networks}
	\small
	{
		\begin{tabular}{ccccccc}
			\toprule
			Networks & Cornell & Texas & Washington & Wisconsin & Cora  & Citeseer \\
			\hline
			$n$     & 195   & 187   & 230   & 265   & 2708  & 3312 \\
			$m$     & 304   & 328   & 446   & 530   & 5429  & 4723 \\
			$K$     & 1703  & 1703  & 1703  & 1703  & 1433  & 3703 \\
			$c$     & 5     & 5     & 5     & 5     & 7     & 6 \\
			\toprule
		\end{tabular}%
	}
	\label{tab1}%
\end{table}%

\begin{table}[ht]
	\centering
	\caption{Comparison of generative models on real networks}
	\resizebox{\textwidth}{!}{
		\begin{tabular}{l c c c c c}
			\toprule
			NMI  & PSB\_PG &  NEMBP & BNPA  & PPSB\_DC  & PCL\_DC \\
			\hline
			Cornell & \textbf{0.3115}$ \pm $0.0576  & 0.1848$ \pm $0.0433 & 0.0777$ \pm $0.0083 & 0.1211$ \pm $0.0231 & 0.0531$ \pm $0.0104 \\
			Texas & \textbf{0.3072}$ \pm $0.0362  & 0.3036$ \pm $0.0252 & 0.2217$ \pm $0.0374 & 0.3056$ \pm $0.0060 & 0.0398$ \pm $0.0067\\
			Washington & \textbf{0.3013}$ \pm $0.0323 & 0.2140$ \pm $0.0378 & 0.2555$ \pm $0.0135 & 0.2391$ \pm $0.0058 & 0.0956$ \pm $0.0181 \\
			Wisconsin & \textbf{0.3729}$ \pm $0.0279 & 0.2863$ \pm $0.0539 & 0.3213$ \pm $0.0102 & 0.2319$ \pm $0.0089 & 0.0463$ \pm $0.0096 \\
			Cora  & 0.3442$ \pm $0.0382  & 0.4166$ \pm $0.0259 & 0.4446$ \pm $0.0169 & \textbf{0.4659}$ \pm $0.0090 & 0.3641$ \pm $0.0228 \\
			Citeseer & 0.2543$ \pm $0.0364 & 0.2298$ \pm $0.0199 & 0.3158$ \pm $0.0077 & \textbf{0.3870}$ \pm $0.0091 & 0.3553$ \pm $0.0335 \\
			\toprule
		\end{tabular}%
	}
	\label{tab2}%
\end{table}%

\begin{figure}[h]
	\centering
	\includegraphics[width=14.5cm]{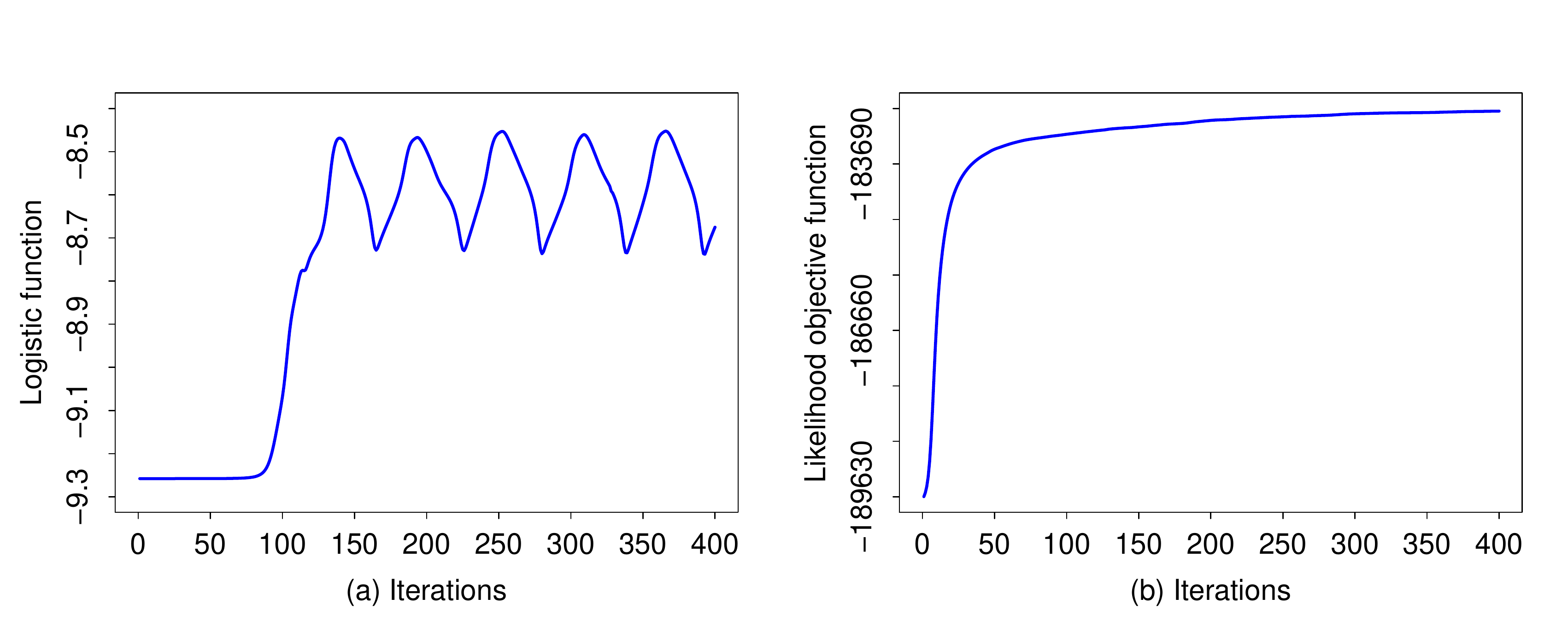}
	\caption{Convergence property of PPSB\_DC (a) and that of PSB\_PG (b) on Cornell network.}\label{fig6}
\end{figure}

From Table \ref{tab2}, it can be seen that the proposed model PSB\_PG has the best accuracy on the first four networks with mixture structures (i.e., Cornell, Texas, Washington, Wisconsin), while PCL\_DC performs the worst on these four network because it is designed to uncover classical community structures. On Cora and Citeseer, PPSB\_DC has showed the best performance. However, PPSB\_DC can not converge in some cases, especially when the initial values of block matrix $\theta$ completely opposite the real structure contained in a given network. For example, Figure \ref{fig6}.a shows a case that the objective function of PPSB\_DC oscillates with the number of iterations. At the same initial values, PSB\_PG shows the tendency of convergence (see Figure \ref{fig6}.b). In summary, PSB\_PG is able to detect general structures, especially good at identifying mixture structures, has flexible interpretability and its convergence is guaranteed by the properties of EM algorithm [36] 

\section{Conclusion and Discussion} 
It is a challenging task on exploring general network structures in attributed networks. In this study, based on the block structure assumption in stochastic blockmodels and the idea of link communities, a principled generative model PSB\_PG is proposed to joint these two types of information. The proposed model gives a unified generative process for generating links and node attributes. The parameters of the model can be easily inferred by expectation-maximization (EM) algorithm with guaranteed convergence. The experimental research has showed that PSB\_PG model not only has the ability to detect a wide range of structures including non-overlapping and overlapping community structures, bipartite structures, mixture structures, etc., but also provides a flexible way to give a semantic interpretation for each detected community, whenever the community contains a topic or multiple topics.

In the future, we intend to further improve the effectiveness of our model by considering degree-corrected parameters into the model akin to PPSB\_DC and NEMBP. Other direct extensions of this work concern more sophisticated inference techniques rather than EM algorithm, such as stochastic variational inference [47], to speed up the computation.

\begin{acknowledgments}
	This work is supported in part by the National Science Foundation of China (granted No. 61473030, No. 61876016 and No. 61632004), the Beijing Municipal Science \& Technology Commission (No. Z181100008918012). The authors thank the anonymous reviewers for their constructive comments.
\end{acknowledgments}

\section*{Appendix A}
From Eq.(5), we know that the log-likelihood function is
$$
\begin{array}{l}
\displaystyle \bar L\left( {D,\Phi ,\Theta } \right) = \frac{1}{2}\sum\limits_{ijrs} {\left[ {a_{ij} q_{ij}^{rs} \log \left( {\frac{{d_{ir} \theta _{rs} d_{js} }}{{q_{ij}^{rs} }}} \right) - d_{ir} \theta _{rs} d_{js} } \right]}  \\ 
\;\;\;\;\;\;\;\;\;\;\;\;\;\;\;\;\;\;\;\;\; + \displaystyle \sum\limits_{ikr} {\left[ {W_{ik} \gamma _{ik}^r \log \left( {\frac{{d_{ir} \phi _{rk} }}{{\gamma _{ik}^r }}} \right) - d_{ir} \phi _{rk} } \right]}. \\ 
\end{array} \eqno(5)
$$
Under the constraint $ \sum\limits_{i = 1}^n {d_{ir} }  = 1 $, and ignoring irrelevant constants, one has
\[
\begin{array}{l}
\tilde L\left( D \right) = \displaystyle \frac{1}{2}\sum\limits_{ijrs} {\left[ {a_{ij} q_{ij}^{rs} \log \left( {d_{ir} d_{js} } \right) - d_{ir} \theta _{rs} d_{js} } \right]}  \\
\;\;\;\;\;\;\;\;\;\;\;\;\; + \displaystyle \sum\limits_{ikr} {\left[ {W_{ik} \gamma _{ik}^r \log \left( {d_{ir} } \right) - d_{ir} \phi _{rk} } \right]}  + \sum\limits_r {\gamma _r \left( {1 - \sum\limits_i {d_{ir} } } \right)}. 
\end{array} \eqno(A1)
\]
Taking the first derivative of the Lagrangian $\tilde L\left( D \right)$ with respect to $d_{ir}$ and set it to be zero, we have
$$
\begin{array}{l}
\displaystyle \frac{{\partial \tilde L\left( D \right)}}{{\partial d_{ir} }} = \frac{{\sum\limits_{js} {\left( {a_{ij} q_{ij}^{rs} } \right)} }}{{d_{ir} }} - \sum\limits_{js} {\left( {\theta _{rs} d_{js} } \right)}  + \frac{{\sum\limits_k {W_{ik} \gamma _{ik}^r } }}{{d_{ir} }} - \sum\limits_k {\left( {\phi _{rk} } \right)}  - \gamma _r  \\ 
\;\;\;\;\;\;\;\;\;\;\;\; = \displaystyle \frac{{\sum\limits_{js} {\left( {a_{ij} q_{ij}^{rs} } \right)} }}{{d_{ir} }} - \sum\limits_s {\theta _{rs} }  + \frac{{\sum\limits_k {W_{ik} \gamma _{ik}^r } }}{{d_{ir} }} - 1 - \gamma _r  = 0. \\ 
\end{array} \eqno(A2)
$$

$$
\sum\limits_{ijs} {\left( {a_{ij} q_{ij}^{rs} } \right)}  - \sum\limits_s {\theta _{rs} }  + \sum\limits_{ik} {\left( {W_{ik} \gamma _{ik}^r } \right)}  - 1 - \gamma _r {\rm{ = }}0. \eqno(A3)
$$
By (A2) and (A3), we can have $d_{ir}$ in Eq.(7) in the following.
$$
d_{ir}  = \frac{{\sum\limits_{js} {a_{ij} q_{ij}^{rs} }  + \sum\limits_k {W_{ik} \gamma _{ik}^r } }}{{\sum\limits_{ijs} {a_{ij} q_{ij}^{rs} }  + \sum\limits_{ik} {W_{ik} \gamma _{ik}^r } }}.
$$

Note that the constraint ${\sum\limits_{r,s = 1}^c {\theta _{rs} } }=1$, one has
$$
\tilde L\left( \Theta  \right) = \frac{1}{2}\sum\limits_{ijrs} {\left[ {a_{ij} q_{ij}^{rs} \log \left( {\theta _{rs} } \right) - d_{ir} \theta _{rs} d_{js} } \right]}  + \lambda \left( {1 - \sum\limits_{r,s = 1}^c {\theta _{rs} } } \right). \eqno(A4)
$$
Taking the first derivative of the Lagrangian $\tilde L\left( \Theta  \right)$ with respect to $\theta _{rs}$ and setting it to be zero, one has
$$
\frac{{\partial \tilde L\left( \Theta  \right)}}{{\partial \theta _{rs} }} = \frac{1}{2}\frac{{\sum\limits_{ij} {\left( {a_{ij} q_{ij}^{rs} } \right)} }}{{\theta _{rs} }} - \frac{1}{2}\sum\limits_{ij} {\left( {d_{ir} d_{js} } \right)}  - \lambda {\rm{ = }}0. \eqno(A5)
$$

$$
\sum\limits_{ijrs} {\left( {a_{ij} q_{ij}^{rs} } \right)}  - 1 - 2\lambda {\rm{ = }}0. \eqno(A6)
$$
By (A5) and (A6), we can derive the equation $\theta _{rs}$ in Eq.(7) below.
$$
\theta _{rs}  = \frac{{\sum\limits_{ij} {a_{ij} q_{ij}^{rs} } }}{{\sum\limits_{ijrs} {a_{ij} q_{ij}^{rs} } }}.
$$

Similarly, for $ {\phi _{rk} } $, one has
$$
\tilde L\left( \Phi  \right) = \sum\limits_{ikr} {\left[ {W_{ik} \gamma _{ik}^r \log \left( {\phi _{rk} } \right) - d_{ir} \phi _{rk} } \right]}  + \sum\limits_r {\eta _r \left( {1 - \sum\limits_{k = 1}^K {\phi _{rk} } } \right)}. \eqno(A7)
$$

$$
\frac{{\sum\limits_i {W_{ik} \gamma _{ik}^r } }}{{\phi _{rk} }} - \sum\limits_i {d_{ir} }  - \eta _r  = 0. \eqno(A8)
$$

$$
\sum\limits_{ik} {W_{ik} \gamma _{ik}^r }  - 1 - \eta _r  = 0. \eqno(A9)
$$
Then, we get the equation $\phi _{rk}$ in Eq.(7) as follows: 
$$
\phi _{rk}  = \frac{{\sum\limits_i {W_{ik} \gamma _{ik}^r } }}{{\sum\limits_{ik} {W_{ik} \gamma _{ik}^r } }}.
$$







\end{document}